\documentclass[prl,twocolumn]{revtex4}
\usepackage{type1cm}
\usepackage[dvipdfm]{graphicx}

\usepackage{amsmath,amssymb,bm,amsthm}

\def\beq{\begin{equation}}
\def\eeq{\end{equation}}
\def\nbeq{\begin{equation*}}
\def\neeq{\end{equation*}}

\def\Vec#1{\bm{#1}}
\def\<{\langle}
\def\>{\rangle}

\renewcommand{\d}{\partial}

\begin{document}
\title{Analysis of Exactness of the Mean-Field Theory in Long-Range Interacting Systems}

\author{Takashi Mori}
\email{
mori@spin.phys.s.u-tokyo.ac.jp}
\affiliation{
Department of Physics, Graduate School of Science,
The University of Tokyo, Bunkyo-ku, Tokyo 113-0033, Japan
}
\affiliation{
CREST, JST, 4-1-8 Honcho Kawaguchi, Saitama, 332-0012, Japan}
\begin{abstract}
Relationships between general long-range interacting classical systems on a lattice and the corresponding mean-field models
(infinitely long-range interacting models) are investigated.
We study systems in arbitrary dimension $d$ for periodic boundary conditions
and focus on the free energy for fixed value of the total magnetization.
As a result, it is shown that the equilibrium free energy of the long-range interacting systems 
are exactly the same as that of the corresponding mean-field models (exactness of the mean-field theory).
Moreover, the mean-field metastable states can be also preserved in general long-range interacting systems.
It is found that in the case that the magnetization is conserved,
the mean-field theory does not give correct property in some parameter region.
\end{abstract}
\maketitle

Long-range interactions appear in several fields~\cite{Lecture_notes2002}:
astrophysics, plasmas, 2D hydrodynamics, and so on.
Recently, it was reported that a model of spin-crossover materials 
has an effective long-range interaction among molecules
due to the coupling to the local lattice distortion~\cite{Miyashita2009}.
It has been shown that statistical mechanics of long-range interacting systems
exhibits several peculiar features:
negative specific heat~\cite{Thirring1970}, long-lived metastable states~\cite{Griffiths1966},
and ensemble inequivalence~\cite{Barre2001}.
As simplified models to study these unfamiliar features, 
the so called mean-field models (infinite-range interacting models) have been often adopted.
From the statistical mechanical point of view, one expects that at least some qualitative natures of systems with slow
decaying long-range interactions are captured by the analysis of corresponding mean-field models~\cite{Lecture_notes2002}.
Indeed, some evidences for the fact that the mean-field model gives an {\it exact} description of equilibrium states
(exactness of the mean-field theory)
in the models with $1/r^{\alpha}$-type long-range interaction
($0\leq\alpha<d$, where $d$ is the spatial dimensionality)
have been obtained by the studies of specific models~\cite{Cannas2000,Tamarit2000,Barre2005,Campa}.
There is a conjecture that the equilibrium properties of long-range interacting spin systems are exactly the same as
those of the corresponding mean-field models~\cite{Cannas2000}.
In this letter, we address the question whether exactness of the mean-field theory holds in general.

We consider a lattice system.
Each lattice site $i$ has a ``spin '' variable $\sigma_i$.
In the Ising model, $\sigma_i=\pm 1$, 
but we do not restrict the variables $\{\sigma_i\}$ to the Ising spins.
They can also take continuous or vector variables.
For example, $\sigma_i=(\cos\theta,\sin\theta)$, $0\le\theta<2\pi$ corresponds to the XY model and
$\sigma_i=\Vec{e}_{\alpha}$ ($\alpha=1,\dots,q$) corresponds to the $q$-state Potts model,
where $\{\Vec{e}_{\alpha}\}$ are unit vectors
satisfying $\Vec{e}_{\alpha}\cdot\Vec{e}_{\beta}=\delta_{\alpha,\beta}$.
We investigate the system of the following type Hamiltonian:
\beq
{\cal H}=-\frac{J}{2}\sum_{ij}K(|\bm{r}_i-\bm{r}_j|)\sigma_i\sigma_j-H\sum_i\sigma_i .
\label{eq:LR_H}
\eeq
Here, $K(|\bm{r}_i-\bm{r}_j|)$ is the interaction potential
and $J$ is the interaction strength.
We consider the ferromagnetic case $J>0$.
The uniform magnetic field is denoted by $H$.
We define the distance $|\bm{r}_i-\bm{r}_j|$ between the lattice points $i$ and $j$
as the shortest distance of these lattice points in periodic boundary conditions.
We consider the following two types of long-range interactions:
the power-law interaction
$K(r)\propto 1/r^{\alpha}$, $0\leq\alpha<d$,
and the Kac interaction~\cite{Kac1963}
$K(r)\propto \gamma^d\phi(\gamma r)$.
Here, $\phi(x)$ is assumed to be non-negative $\phi(x)\geq 0$ and integrable
$\int d^dx\phi(x)<+\infty$.
Moreover, we assume that there is a positive and decreasing function $\psi(x)$ such that
$|\phi'(x)|\leq\psi(x)$ and
$\int d^dx\psi(x)<+\infty$.
These assumptions are necessary to justify the coarse-graining of the Hamiltonian (to be explained below).
A typical example of the Kac interaction is the exponential form,
$K(r)\propto \gamma^de^{-\gamma r}$.
In this case, $\phi(x)=\psi(x)=e^{-x}$.

We take the limit $\gamma\rightarrow 0$ in the Kac interaction.
We consider the two limiting procedures:
the van der Waals limit, $\gamma\rightarrow 0$ {\it after} $L\rightarrow\infty$~\cite{Lebowitz_Penrose1966}
and the long-range limit, $\gamma\rightarrow 0$ with $\gamma L=\text{const.}$
The former limit corresponds to the situation that the interaction range $\gamma^{-1}$ is much longer than
the microscopic length scale (the lattice interval) but much shorter than the system size $L$.
The latter limit corresponds to the situation that the interaction range is comparable with the system size.
These two limits lead to different kind of behavior.

Note that the thermodynamic limit does not exist for long-range interactions in the usual sense.
To restore the thermodynamic limit, we adopt the Kac prescription~\cite{Kac1963},
$\sum_{\bm{r}_i\neq 0}K(|\bm{r}_i|)=1$,
where the interaction is normalized and depends on the system size.

We focus on the free energy restricted to a fixed magnetization $m$,
\beq
\exp(-\beta F(m,T,H))=
\sum_{\{\sigma_i\}}\delta\left(\frac{1}{N}\sum_i\sigma_i,m\right)e^{-\beta {\cal H}},
\label{eq:LR_free}
\eeq
which corresponds to the Landau free energy, and we call $F(m,T,H)$ merely ``free energy'' hereafter.
The parameter $\beta=1/T$ is the inverse temperature and we set the Boltzmann constant to unity.
The symbol $\delta(a,b)$ is the Kronecker delta.
For {\it conserved systems} (i.e. systems whose magnetization is fixed),
$m$ is a fixed parameter and the term $-Hm$ is just a constant.
Then, $F(m,T,H)$ gives the equilibrium free energy.

In {\it non-conserved systems} (i.e. systems whose magnetization is not fixed), the equilibrium free energy is given by
the minimum value of $F(m,T,H)$ with respect to $m$.
Local but not global minima are interpreted as metastable states.
In preceding works, exactness of the mean-field theory is investigated only for non-conserved systems
~\cite{Cannas2000,Tamarit2000,Barre2005,Campa}.
The merit to consider $F(m,T,H)$ is that we can treat both conserved and non-conserved systems.
It should be noted that the ensemble with fixed magnetization and that with fixed magnetic field are essentially different
in long-range interacting systems.
Moreover, it allows us to discuss metastable states as local minimum points of the free energy.

In the long-range interacting systems, one expects that 
only long wavelength modes play important roles for macroscopic behavior.
In fact, it is possible to perform coarse-graining {\it exactly} for long-range interacting models~\cite{Barre2005}.
Let us divide the lattice system into blocks of the linear dimension $l$. 
The number of blocks is $(L/l)^d$, where $L$ denotes the system size, and each block has $l^d$ sites.
We introduce a local coarse-grained variable $m_k$ as 
$m_k=\frac{1}{l^d}\sum_{i\in B_k}\sigma_i$
in each block $B_k$, $k=1,2,\dots ,(L/l)^d$.
We define the position $\bm{x}_k=\bm{r}_k/L$, where $\bm{r}_k$ is the central position of a block $B_k$.
We also define $m(\bm{x}_k)\equiv m_k$.
We take the limit $L\rightarrow\infty$, $l\rightarrow\infty$ with $l/L\rightarrow 0$ (continuous limit).
In this limit, $\bm{x}_k$ becomes a continuous variable $\bm{x}$.
For long-range interacting models, in a proper choice of $l(L)$ as a function of the system size $L$
such that $\lim_{L\rightarrow\infty}l(L)/L=0$,
the Hamiltonian is written only by $m(\bm{x})$ in the thermodynamic limit~\cite{Barre2005}:
${\cal H}=\bar{{\cal H}}[m(\bm{x})]+o(N)$,
where
\begin{align}
\bar{{\cal H}}[m(\bm{x})]
=&-\frac{NJ}{2}\int_{0}^{1}d^dx\int_{0}^{1}d^dy U(\bm{x}-\bm{y})m(\bm{x})m(\bm{y})
\nonumber \\
&-NH\int_{0}^{1}d^dx m(\bm{x}).
\label{eq:LR_cgH}
\end{align}
Here the interaction potential $U(x)$ is given by
\beq
U(\bm{x})=\lim_{L\rightarrow\infty}L^dK(Lx).
\label{eq:LR_cg_potential}
\eeq
The integration in~(\ref{eq:LR_cgH}) means
$\int_{a}^bd^dx\equiv \int_{a}^bdx_1\dots\int_{a}^bdx_d.$
The fact that the Hamiltonian is written only by the coarse-grained magnetization $m(\bm{x})$ means that
for a fixed system size $L$ (it is very large), there is some length scale $l(L)$ in which the magnetization is uniform,
though the whole system may be inhomogeneous.
For example, in the case of the power-law interaction $K(r)\sim 1/r^{\alpha}$,
this length scale is given by $l(L)\sim L^{1-\alpha/d}$.
While the size of a block $l(L)$ is itself macroscopic when $L\rightarrow\infty$,
the blocks are not independent each other.
It means that we cannot divide a macroscopic system into two macroscopic subsystems
without any macroscopic change.

Performing the Fourier expansion in Eq.~(\ref{eq:LR_cgH}), we obtain the following expression:
\beq
\bar{{\cal H}}=-\frac{NJ}{2}\sum_{\bm{n}}U_{\bm{n}}|\hat{m}_{\bm{n}}|^2-NH\hat{m}_0,
\label{eq:fourier}
\eeq
where
$m(\bm{x})=\sum_{\bm{n}}\hat{m}_{\bm{n}}e^{2\pi i\bm{n}\cdot\bm{x}}$
and
$U_{\bm{n}}=\int_{0}^{1}d^dx U(\bm{x})\cos(2\pi\bm{n}\cdot\bm{x})$
with $\bm{n}\in\bm{Z}^d$.
Here, the Kac prescription implies that
$U_0=\int_{0}^{1}d^dx U(\bm{x})=1$.
We call $\{U_{\bm{n}}\}$ the interaction eigenvalues.
Dividing the Hamiltonian into two parts: the mode of $\bm{n}=0$ and the others, we have
\beq
\bar{{\cal H}}=\left(-\frac{NJ}{2}\hat{m}_0^2-NH\hat{m}_0\right)
-\frac{NJ}{2}\sum_{\bm{n}\neq 0}U_{\bm{n}}|\hat{m}_{\bm{n}}|^2.
\eeq
We call the first term of the above equation the mean-field model corresponding to (\ref{eq:LR_H}).

The form of the coarse-grained potential $U(\bm{x})$ is given by
$U(\bm{x})=A/x^{\alpha}$ for the power-law interaction,
$U(\bm{x})=\delta(\bm{x})$ for the Kac interaction with the van der Waals limit, and
$U(\bm{x})=B\gamma_0^d\phi(\gamma_0 x)$ for the Kac interaction with the long-range limit.
The constants $A$ and $B$ are determined by the normalization condition $U_0=1$.
Note that $U_{\bm n}=1$ for any $\bm{n}$ for the Kac interaction with the van der Waals limit.
It is known that the free energy is given by the mean-field theory 
with the Maxwell construction in the van der Waals limit~\cite{Lebowitz_Penrose1966}.
Therefore, the free energy does not depend on its detailed interaction form in the van der Waals limit.
On the other hand, for the long-range limit or the power-law interaction, 
whose interaction range is comparable with the system size,
the free energy does depend on the interaction form, as we see below.
The modes with $0<U_{\bm n}<1$ play an important role.


Now, we evaluate the free energy $F(m,T,H)$.
Since the coarse-graining can be exactly performed,
the Hamiltonian $\cal{H}$ can be replaced by $\bar{{\cal H}}[m(\bm{x})]$.
The summation of (\ref{eq:LR_free}) is divided into two parts,
$\int_{\hat{m}_0=m}{\cal D}m(\bm{x})$ and
$\sum_{\{\sigma_i\} \text{ with fixed $m(\bm{x})$}}$, namely,
the profile of $m(\bm{x})$ and the configurations inside the blocks.
Note that the Hamiltonian depends only on $m(\bm{x})$.
Writing the number of states with the fixed magnetization $m$ by $W=\exp(S(m))$, where $S(m)$ is the {\it entropy},
we have
$\sum_{\{\sigma_i\} \text{ with fixed $m(\bm{x})$}}1
=\exp\left(\int_0^1S(m(\bm{x}))d^dx\right)$.
Therefore, we obtain
\beq
e^{-\beta F(m,T,H)}
=\int_{\hat{m}_0=m}{\cal D}m(\bm{x})
e^{-\beta {\cal F}[m(\bm{x}),T,H]},
\eeq
where ${\cal F}[m(\bm{x}),T,H]\equiv \bar{\cal H}[m(\bm{x})]-T\int_0^1S(m(\bm{x}))d^dx$
is called the free energy functional.
By using the saddle-point method, the free energy is given by the following minimization problem:
\beq
F(m,T,H)=\min_{\{m(\bm{x})|\hat{m}_0=m\}}{\cal F}[m(\bm{x}),T,H].
\label{eq:functional}
\eeq
The free energy functional is the same as the mean-field free energy $F_{\rm MF}$ when $m(\bm{x})=m$,
therefore, the upper-bound of the free energy is obtained,
$F(m,T,H)\leq F_{\rm MF}(m,T,H)$.
We also found that $F(m,T,H)<F_{\rm MF}(m,T,H)$ when $\d^2F_{\rm MF}(m,T/U_{\rm max},H)/\d m^2<0$,
that is, $F(m,T,H)$ does not agree with $F_{\rm MF}(m,T,H)$.
It is because the uniform magnetization profile $m(\bm{x})=m$ 
gives the local maximum value of the free energy functional when $\d^2F_{\rm MF}(m,T/U_{\rm max},H)/\d m^2<0$.
Here, $U_{\rm max}\leq 1$ is the largest interaction eigenvalue except for $U_0=1$.

Taking into account of the relations
$\sum_{\bm{n}\neq 0}U_{\bm{n}}|\hat{m}_{\bm{n}}|^2
\leq U_{\rm max}\sum_{\bm{n}\neq 0}|\hat{m}_{\bm{n}}|^2$
and
$\int_0^1m(\bm{x})^2d^dx=m^2+\sum_{\bm{n}\neq 0}|\hat{m}_{\bm{n}}|^2$,
the free energy functional is evaluated as
\begin{align}
{\cal F}[m(\bm{x}),T,H]
\geq &F_{\rm MF}(m,T,H)-U_{\rm max}\bigg[F_{\rm MF}(m,T_{\rm eff},H) \nonumber \\
&-\int_0^1d^dx F_{\rm MF}(m(\bm{x}),T_{\rm eff},H)\bigg],
\label{eq:LR_inequality}
\end{align}
where $T_{\rm eff}\equiv T/U_{\rm max}$.
Here the following equality holds:
\begin{align}
\min_{\{m(\bm{x})|\hat{m}_0=m\}}\left[\int_0^1d^dx F_{\rm MF}(m(\bm{x}),T_{\rm eff},H)\right]
\nonumber \\
={\rm CE}\{ F_{\rm MF}(m,T_{\rm eff},H)\},
\label{eq:LR_CE}
\end{align}
where CE means the convex envelope.
The convex envelope of a function $g(x)$ is defined as
the maximum convex function not exceeding $g(x)$.
Thus, it is concluded that
$F(m,T,H)\geq F_{\rm MF}(m,T,H)
-U_{\rm max}\Delta F_{\rm MF}(m,T_{\rm eff},H)$,
where we define $\Delta F_{\rm MF}\equiv F_{\rm MF}-{\rm CE}\{F_{\rm MF}\}$. 
It gives the lower bound of the free energy.
Finally, we obtain
\begin{align}
&F_{\rm MF}(m,T,H)-U_{\rm max}\Delta F_{\rm MF}(m,T/U_{\rm max},H)
\nonumber \\
&\leq F(m,T,H)\leq F_{\rm MF}(m,T,H).
\label{eq:main}
\end{align}
Only the inequality $U_{\bm{n}}\leq U_{\rm max}$ is used to derive the lower bound.
In the Kac interaction with the van der Waals limit, $U_{\bm n}=1$ for any $\bm{n}$,
and hence the lower bound in Eq.~(\ref{eq:main}) is realized and thus,
$F(m,T,H)={\rm CE}\{F_{\rm MF}(m,T,H)\}$ holds.
This result is consistent with~\cite{Lebowitz_Penrose1966}.

According to Eq.~(\ref{eq:main}),
the parameter region $(m,T,H)$ is classified into the following three regions:
\begin{description}
\item[Region A:] the region where $\Delta F_{\rm MF}\left(m,\frac{T}{U_{\rm max}},H\right)=0$.
In this region, the mean-field model gives the exact free energy, $F(m,T,H)=F_{\rm MF}(m,T,H)$.
\item[Region B:] the region where $\Delta F_{\rm MF}\left(m,\frac{T}{U_{\rm max}},H\right)>0$ and 
$\frac{\d^2}{\d m^2}F_{\rm MF}\left(m,\frac{T}{U_{\rm max}},H\right)\geq 0$.
In this region, the uniform configuration is locally stable but not necesarily globally stable and $F(m,T,H)\leq F_{\rm MF}(m,T,H)$.
\item[Region C:] the region where $\frac{\d^2}{\d m^2}F_{\rm MF}\left(m,\frac{T}{U_{\rm max}},H\right)< 0$.
In this region, the uniform configuration is unstable and 
the free energy is not given by the mean-field free energy, $F(m,T,H)< F_{\rm MF}(m,T,H)$.
\end{description}

\begin{figure}[t]
\begin{center}
\includegraphics[clip,width=7cm]{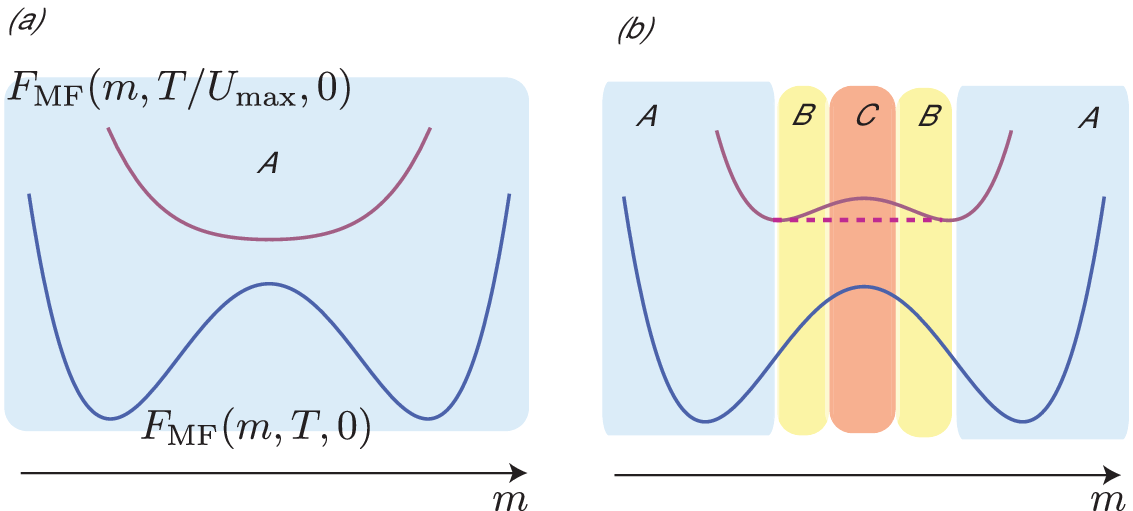}
\caption{(Color online)
An illustrative explanation of the regions A, B, and C in the Ising model at $H=0$.
The blue lines represent the mean-field free energy at the temperature $T$.
The red lines represent the mean-field free energy 
at the effective temperature $T/U_{\rm max}$.
(a) The case of $T/U_{\rm max}>T_{\rm C}>T$.
(b) The case of $T_{\rm C}>T/U_{\rm max}>T$.}
\label{fig:LR_MF_region}
\end{center}
\end{figure}

Notice that this classification is determined only by $F_{\rm MF}$ and $U_{\rm max}$.
We can calculate both $F_{\rm MF}$ and $U_{\rm max}$ for specific models.
Hence, we can specify these three regions explicitly.
In Fig.~\ref{fig:LR_MF_region}, a typical example of
the regions A, B, and C are depicted.
In the case $T_{\rm eff}>T_{\rm C}>T$ where $T_{\rm C}$ is the mean-field critical temperature (Fig.~\ref{fig:LR_MF_region} (a)), 
$F_{\rm MF}(m,T_{\rm eff},H)$ is a convex function, and thus the whole $m$ is in the region A.
On the other hand, in the case $T_{\rm C}>T_{\rm eff}>T$ (Fig.~\ref{fig:LR_MF_region} (b)), 
the region where $F_{\rm MF}(m,T_{\rm eff},H)$ is concave exists, and it is the region C.
The region between A and C is the region B.
In the region A and a part of B, $F(m,T,H)=F_{\rm MF}(m,T,H)$ holds.
We call this region ``the MF region''.
On the other hand, in the region C and the other part of B, $F(m,T,H)\neq F_{\rm MF}(m,T,H)$.
This region is ``the non-MF region''.

\begin{figure}[t]
\begin{center}
\begin{tabular}{cc}
(a)&(b) \\
\includegraphics[clip,width=2cm,angle=-90]{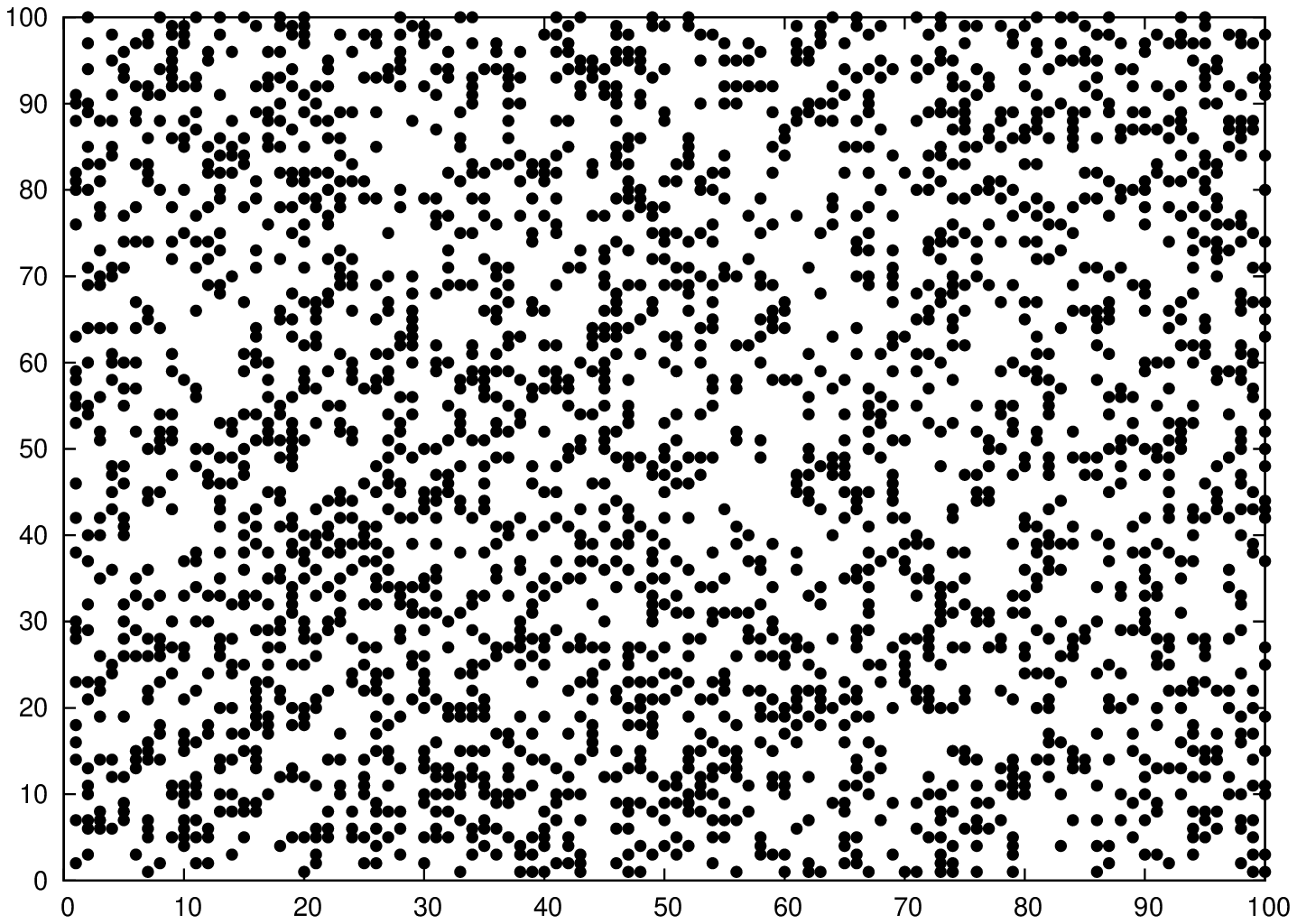}&
\includegraphics[clip,width=2cm,angle=-90]{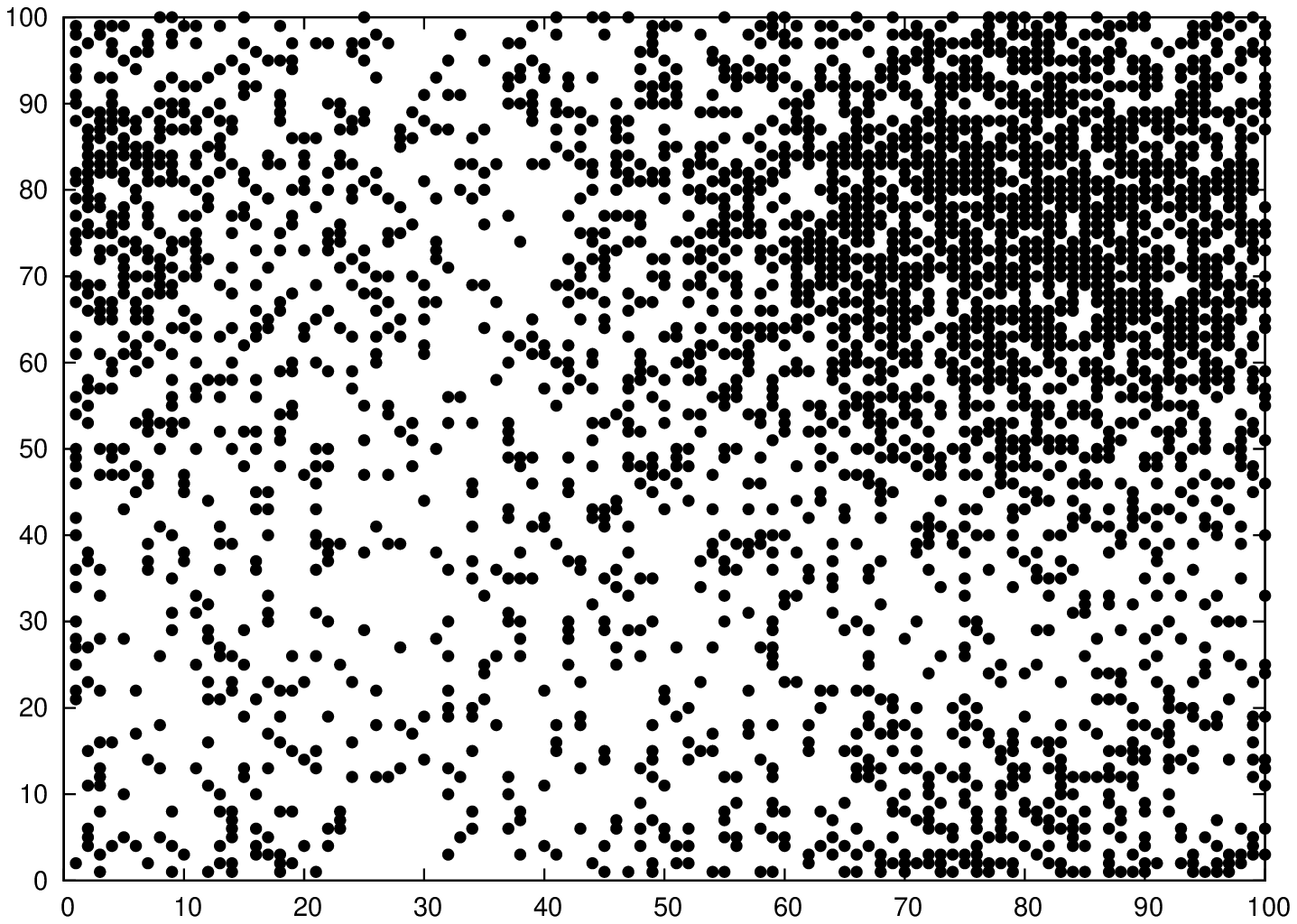}
\end{tabular}
\caption{Typical equilibrium configurations in the 2D lattice gas model (conserved Ising model)
 with $1/r$ type long-range interaction.
 The black points correspond to the occupied state $\sigma_i=+1$.
 The parameters are set to be $m=0.5$, $J=2$ and $L=100$.
(a) The case of $T=0.6$.
These parameters belong to the MF region.
(b) The case of $T=0.4$.
These parameters belong to the non-MF region.}
\label{fig:LR_eq_cluster}
\end{center}
\end{figure}

In conserved systems,
$F_{\rm eq}(m,T)=F(m,T,H=0)$ and
the equilibrium property of the long-range interacting system is {\it exactly} the same 
as that of the corresponding mean-field model in the MF region,
but they are not the same in the non-MF region.
In the MF region, the magnetization is uniform (see Fig.~\ref{fig:LR_eq_cluster}~(a)).
In the non-MF region, inhomogeneity appears, which is demonstrated in Fig.~\ref{fig:LR_eq_cluster}~(b),
because the deviation from the mean-field model is 
due to the non-zero wavenumber components $\hat{m}_{\bm{n}\neq 0}$ of the magnetization.
In fact, a kind of phase transition occurs in the boundary of the MF and the non-MF region,
and the cluster is formed at equilibrium in the non-MF region.
However, this clustering should not be understood by the phase separation.
In long-range interacting systems, a part of the system is not independent of the other part of the system,
and we cannot separate the system into two parts without any macroscopic change.
Anyway, in conserved systems, the long-range interacting model is {\it not} fully describable by the mean-field model
even in equilibrium.

\begin{figure}[t]
\begin{center}
\begin{tabular}{cc}
(a)&(b) \\
\includegraphics[clip,width=4.3cm]{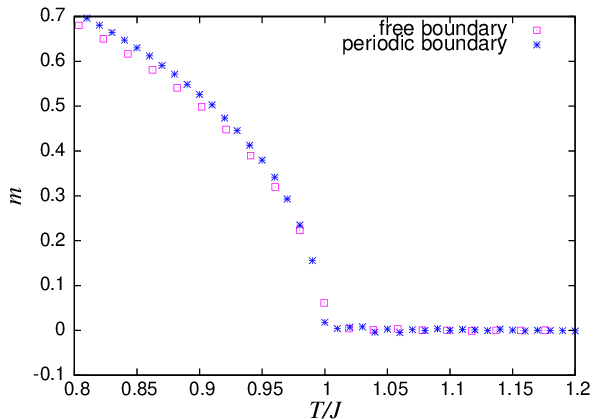}&
\includegraphics[clip,width=4.3cm]{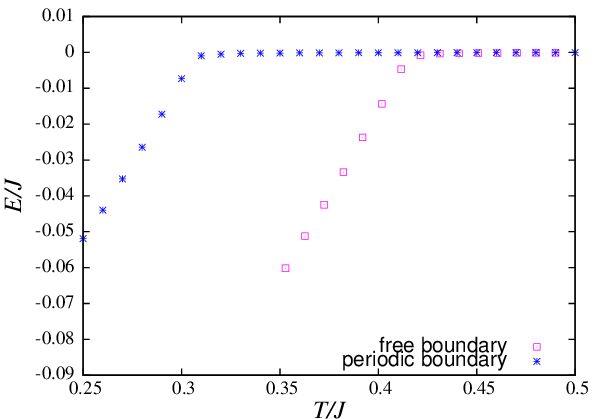}
\end{tabular}
\caption{The comparison between periodic and free boundary conditions by the Monte Carlo calculations.
The system size is $L=200$.
(a) Equilibrium magnetizations in non-conserved systems.
(b) Equilibrium internal energies in conserved systems at $m=0$.}
\label{fig:LR_pbc}
\end{center}
\end{figure} 

In non-conserved systems, the equilibrium magnetization $m_{\rm eq}$ is determined by 
$\min_m[F(m,T,H)]=F(m_{\rm eq},T,H)$.
Because $\Delta F_{\rm MF}(m_{\rm eq},T_{\rm eff},H)=0$ and all the equilibrium states belong to the region A,
it is concluded that the equilibrium property is {\it always equivalent} to the mean-field model,
which was conjectured by Cannas et al.~\cite{Cannas2000}.
Moreover, our result implies that many (but not all) mean-field metastable states are also preserved 
in general long-range interacting systems
because local minima of the free energy are in the region A or B.
Indeed, we have checked by numerical calculations that in the Glauber dynamics, 
relaxation times from the local minima of the free energy
to the equilibrium states are identical to those of the corresponding mean-field model (not shown).
To study some universal dynamical aspects is an important future problem.


Finally, we comment on boundary conditions.
We imposed the periodic ones in this letter, but boundary effects will be important for long-range interacting systems.
Let us redefine the interaction eigenvalues for free boundary.
In this case, the interaction eigenvalues are not given by the Fourier coefficients.
In general, we regard $U_{ij}=K(|\bm{r}_i-\bm{r}_j|)$ as a matrix and denote its eigenvalues by $\{u_k\}$
($u_0\geq u_1\geq u_2\geq\dots$). 
We call $u_k$ as an interaction eigenvalue and normalize the interaction to set $u_0=1$.
Regarding $u_1$ as $U_{\rm max}$, we have found by Monte Carlo calculations
that the above redefinition of interaction eigenvalues  allow us to extend our result to the free boundary case.
For example, see Fig.~\ref{fig:LR_pbc}.
We consider the two dimensional Ising model ($\sigma=\pm 1$) on the square lattice with the $1/r$ type interaction.
In this case, $U_{\rm max}=0.31$ for periodic boundary and $U_{\rm max}=0.42$ for free boundary.
In Fig.~\ref{fig:LR_pbc} (a), the equilibrium magnetizations in non-conserved systems are depicted.
Our result states that the mean-field theory is exact and the phase transition occurs at $T/J=1$.
In Fig.~\ref{fig:LR_pbc} (b), the equilibrium energies in conserved systems (we set $m=0$) are depicted.
Our result states that when $T/J>U_{\rm max}$ (the region A), the system is uniform and the energy is constant,
but when $T/J<U_{\rm max}$ (the region C), the inhomogeneity grows.
These numerical results are consistent with our result for both boundary conditions.
The boundary effects will be examined in more detail elsewhere.

In conclusion, we have studied the robustness of the results of the mean-field model against the interaction forms.
We revealed that
the results of the mean-field model are fully reliable in a wide parameter region (the MF region),
but there is the other region (the non-MF region) where the mean-field model cannot describe long-range interacting systems.
Properties of states in the non-MF region are not obvious yet.
It is a future problem to study them.

The author thank Professor Seiji Miyashita for valuable comments and discussions.
This work was supported by JSPS Research Fellowship for Young Scientists.


\end{document}